\documentclass{ws-ijmpcs}

\begin{document}

\markboth{K. B. Boshkayev et al.}
{Equilibrium structure of white dwarfs at finite temperatures}

\catchline{}{}{}{}{}
%
%%%%%%%%%%%%%%%%%%%%%%%%%%%%%%%%%%%%%%%%%%%%%%%%%%%%%%%%%%%%%%%%%%%%

\title{Equilibrium structure of white dwarfs at finite temperatures}

\author{K. A. Boshkayev,$^{1,2,^*}$ J. A. Rueda,$^{2}$ B. A. Zhami,$^{1}$ \\ Zh. A. Kalymova$^{1}$ and G. Sh. Balgymbekov$^{1}$}

\address{$^1$Department of Physics and Technology, Al-Farabi Kazakh National University,\\
Al-Farabi avenue 71, Almaty, 050040, Kazakhstan\\
$^2$International Center for Relativistic Astrophysics Network,\\
Piazza della Repubblica 10, Pescara, I-65122, Italy\\
$^*$kuantay@mail.ru}
\maketitle
%
%\begin{history}
%\received{Day Month Year}
%\revised{Day Month Year}
%\published{Day Month Year}
%\end{history}
%
\begin{abstract}\label{sec:1}Recently, it has been shown by S.~M. de Carvalho et al. (2014) that the deviations between the degenerate case and observations were already evident for 0.7-0.8 M$_{\odot}$ white dwarfs. Such deviations were related to the neglected effects of finite temperatures on the structure of a white dwarf. Therefore, in this work by employing the Chandrasekhar equation of state taking into account the effects of temperature we show how the total pressure of the white dwarf matter depends on the mass density at different temperatures. Afterwards we construct equilibrium configurations of white dwarfs at finite temperatures. We obtain the mass-radius relations of white dwarfs for different temperatures by solving the Tolman-Oppenheimer-Volkoff equation, and compare them with the estimated masses and radii inferred from the Sloan Digital Sky Survey Data Release 4. 
\end{abstract}

\ccode{PACS numbers: 03.75.Ss; 04.40.Dg; 95.30.Sf; 97.20.Rp}
%%%%%%%%%%%%%%%%%%%%%%%%%%%%%%%%%%%%%%%%%%%%%%%%%%%%%%%%%%%%%%%%%%%%%%%%%%%%%%%%%%%%%%%%%%%%%%%%%%%%%%%%%%%%%%%%%%%%%%%%%%%%%%%%%%%%%%%%%%%%%%%%
\section{Introduction}\label{sec:2}	

In astrophysics it is of high importance to construct a realistic physical model of stellar compact objects such as neutron stars and white dwarfs (WDs) which fits with observations. Consequently, all physical phenomena and quantities must be duly taken into account in the equation of state (EoS). Up to now there exist three EoS to describe the degenerate matter of WDs: the Chandrasekhar EoS,\cite{chandra1} the Salpeter EoS\cite{salp1,salp2} and the Relativistic Feynman-Metropolis-Teller (RFMT) EoS.\cite{FMT1,rotondo,rotondo2} The main differences, advantages and drawbacks among theses EoS are discussed by Rotondo et al\cite{rotondo2}. in detail. Moreover, according to Shapiro et al.\cite{rotondo2,shapiro1} it is necessary to investigate WDs in general relativity (GR) in order to analyze their stability though the corrections of GR can be neglected for low mass WDs.

In this work, we investigate hot white dwarfs by employing the Chandrasekhar EoS\cite{chandra1} including the finite-temperature effects. Similar approach to include the effects of finite temperatures in case of the RFMT EoS has been used by S.~M. de Carvalho et al.\cite{sheyse}

Here we perform similar analyses following S.~M. de Carvalho et al.\cite{sheyse} in order to construct the mass-radius relations of white dwarfs at finite temperatures in GR for the sake of completeness. We compare our results with the estimated masses and radii from the Sloan Digital Sky Survey Data Release 4 (SDSS-E06 catalog).\cite{tremblay} 
%%%%%%%%%%%%%%%%%%%%%%%%%%%%%%%%%%%%%%%%%%%%%%%%%%%%%%%%%%%%%%%%%%%%%%%%%%%%%%%%%%%%%%%%%%%%%%%%%%%%%%%%%%%%%%%%%%%%%%%%%%%%%%%%%%%%%%%%%%%%%%%%

\section{Equation of State}\label{sec:4}
For the sake of clarity we make use of the Chandrasekhar EoS since it is the simplest EoS for WD matter, is well known and widely used in the description of physical properties of WDs.\cite{rotondo2}

\subsection{Chandrasekhar EoS at finite temperatures}\label{sec:42}

In general, the EoS is determined through the total pressure and the total energy density of matter. Within the Chandrasekhar's approximation, the total pressure is due to the pressure of electrons ${P}_{e}$ since the pressure of positive ions ${P}_{N}$ (naked nuclei) is insignificant whereas the energy density is due to the energy density of nuclei ${\cal E}_{N}$ as the energy density of the degenerate electrons ${\cal E}_{e}$ is negligibly small. Thus, the Chandrasekhar EoS is given by

\begin{equation}\label{eq:ECh0}
{\cal E}_{Ch}={\cal E}_{N}+{\cal E}_{e}\approx {\cal E}_{N}=\frac{A}{Z} M_{u}c^{2}n_{e},
\end{equation}
\begin{equation}\label{eq:PCh0}
{P}_{Ch}={P}_{N}+{P}_{e}\approx{P}_{e},
\end{equation}
where $A$ is the average atomic weight, $Z$ is the number of protons, $M_{u}=1.6604\times10^{-24}$ g is the unified atomic mass, $c$ is the speed of light and $n_{e}$ is the electron number density.
In general, the electron number density follows from the Fermi-Dirac statistics and is determined by\cite{landau,arnett1}
\begin{equation}\label{eq:ne}
n_e=\frac{2}{(2\pi \hbar)^{3}} \int_0^\infty \frac{4\pi p^{2} dp}
{\exp\left[\frac{\tilde{E}(p)- \tilde{\mu}_{e}(p)}{k_B T}\right] +1} ,
\end{equation}
where $k_B$ is the Boltzmann constant, $\tilde{\mu}_{e}$ is the electron chemical potential without the rest-mass, and $\tilde{E}(p)= \sqrt{c^2 p^2+m^2_e c^4} - m_{e}c^2$, with $p$ and $m_e$ the electron momentum and rest-mass, respectively.

It is possible to show that (\ref{eq:ne}) can be written in an alternative form as
\begin{equation}
n_{e}=\frac{8\pi \sqrt{2}}{(2 \pi \hbar)^3} m^3 c^3 \beta^{3/2} \left[ F_{1/2} (\eta,\beta) + \beta F_{3/2} (\eta,\beta) \right],
\label{eq:ne2}
\end{equation}
where 
\begin{equation}
F_{k} (\eta,\beta)=\int_{0} ^{\infty} \frac{t^k \sqrt{1+(\beta/2)t}}{1+ e^{t-\eta}}\, dt
\label{eq:Fk1}
\end{equation}
is the relativistic Fermi-Dirac integral, $\eta=\tilde{\mu}_e/(k_B T)$, $t=\tilde{E}(p)/(k_{B}T)$ and $\beta=k_{B}T/(m_e c^2)$ are degeneracy parameters.\cite{sheyse,arnett1}

Consequently, the total electron pressure for $T\neq0$ K is given by 
\begin{align}\label{eq:Pe2}
P_e=\frac{2^{3/2}}{3 \pi^2 \hbar^3} m_{e}^4 c^5 \beta^{5/2} \left[ F_{3/2} (\eta,\beta)\right.+ \left. \frac{\beta}{2} F_{5/2} (\eta,\beta) \right].
\end{align}

\begin{figure}[t]
%\vspace*{-1cm}
\centerline{\includegraphics[scale=0.58]{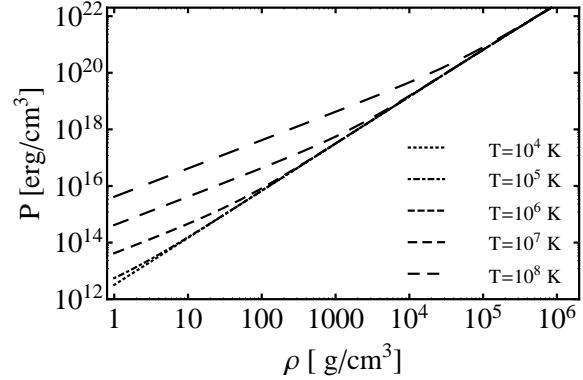}}
%\vspace*{0.2cm}
\caption{Total pressure as a function of the mass density in the case of $\mu=A/Z=2$ white dwarf for selected temperatures in the range $T=(10^4-10^8)$ K.}
\label{fig:f1}
\end{figure}

In Fig.~\ref{fig:f1} we plot the total pressure (\ref{eq:Pe2}) as a function of the total density (\ref{eq:ECh0}) of the system for selected temperatures $T=\left[10^4,10^5,10^6,10^7,10^8\right]$ K. As one can see the effects of temperature is essential in the range of small densities.
%%%%%%%%%%%%%%%%%%%%%%%%%%%%%%%%%%%%%%%%%%%%%%%%%%%%%%%%%%%%%%%%%%%%%%%%%%%%%%%%%%%%%%%%%%%%%%%%%%%%%%%%%%%%%%%%%%%%%%%%%%%%%%%%%%%%%%%%%%%%%%%%
\subsection{Chandrasekhar EoS at T=0}\label{sec:41}

When $T=0$ one can write for the number density of the degenerate electron gas the following expression from the Eq.~(\ref{eq:ne}) 
\begin{equation}
n_e=\int_{0}^{P^F_e}\frac{2}{(2\pi\hbar)^3}d^{3}p=\frac{8 \pi}{(2\pi\hbar)^3}\int_{0}^{P^F_e}p^2dp=\frac{(P^F_e)^3}{3{\pi}^2{\hbar}^3}= \frac{(m_e c)^3}{3{\pi}^2{\hbar}^3}x_e^3
\end{equation}
and the total electron pressure
\begin{eqnarray}
P_e &=& \frac{1}{3} \frac{2}{(2 \pi \hbar)^3} \int_0^{P^F_e} \frac{c^2 p^2}{\sqrt{c^2 p^2+m^2_e c^4}} 4 \pi p^2 dp \nonumber \\
&=& \frac{m^4_e c^5}{8 \pi^2 \hbar^3}[x_e \sqrt{1+x^2_e}(2x^2_e/3-1)+{\rm arcsinh}(x_e)] \label{eq:eos2}\, ,
\end{eqnarray}
where $x_e = P^F_e/(m_e c)$ is the dimensionless Fermi momentum.

%%%%%%%%%%%%%%%%%%%%%%%%%%%%%%%%%%%%%%%%%%%%%%%%%%%%%%%%%%%%%%%%%%%%%%%%%%%%%%%%%%%%%%%%%%%%%%%%%%%%%%%%%%%%%%%%%%%%%%%%%%%%%%%%%%%%%%%%%%%%%%%%

\section{Equations of Stellar Structure and Equilibrium}\label{sec:5}
From spherically symmetric metric
\begin{equation}\label{eq:metric}
ds^2 = e^{\nu(r)} c^2 dt^2 - e^{\lambda(r)}dr^2 - r^2 d\theta^2 - r^2 \sin^2 \theta d\varphi^2\, ,
\end{equation}
the equations of equilibrium can be written in the Tolman-Oppenheimer-Volkoff (TOV) form,
\begin{eqnarray}
\frac{d \nu(r)}{dr} &=& \frac{2 G}{c^2} \frac{4 \pi r^3 P(r)/c^2 + M(r)}{r^2 \left[1 - \frac{2 G M(r)}{c^2 r}\right]}\,\label{eq:nu},\\
\frac{d M(r)}{dr} &=& 4 \pi r^2 \frac{{\cal E}(r)}{c^2}\,\label{eq:mr},\\
\frac{d P(r)}{dr} &=& - \frac{1}{2} \frac{d \nu(r)}{dr} [{\cal E}(r)+P(r)]\,\label{eq:pr},
\end{eqnarray}
where the mass enclosed at the distance $r$ is determined through $e^{-\lambda(r)} = 1 - 2 G M(r)/(c^2 r)$, ${\cal E}(r)=c^2 \rho(r)$ is the total energy density, and $P(r)$ is the total pressure, given by~Eqs.~(\ref{eq:ECh0}) and (\ref{eq:PCh0}).

From the Eqs. (\ref{eq:nu}) and (\ref{eq:pr}) the total pressure can be rewritten in the following form
\begin{equation}
\frac{dP(r)}{dr}=-\frac{G M(r) \rho (r)}{r^2}\left[1+\frac{P(r)}{\rho (r) c^2}\right]\left[1+\frac{4 \pi
    r^2 P(r)}{M(r) c^2}\right]\left[1-\frac{2 G M(r)}{r c^2}\right]^{-1}.
\end{equation}

The TOV equation completely determines the structure of a spherically symmetric body of isotropic material in equilibrium. The first two factors in square brackets represent special relativistic corrections of order $1/c^2$ that arise from the mass-energy relation so that the denominators, $\cal E$ and $Mc^2$, vary relativistically in connection with Einstein's famous equation $E=mc^2$. The last term in brackets is a general relativistic correction due to non-negligible strength of the gravitational potential (in units of $c^2$) and the meaning of $M(r)$ as the total integrated mass out to a radial distance $r$. These corrections each act to strengthen the gravitational interaction. If terms of order $1/c^2$ are neglected, the TOV equation becomes the Newtonian hydrostatic equation, 

\begin{equation}
\frac{dP(r)}{dr}=-\frac{G M(r) \rho (r)}{r^2},
\label{eq:ness1}
\end{equation}
\begin{equation}
\frac{dM(r)}{dr}=4 \pi  r^2 \rho (r)
\label{eq:ness2}
\end{equation}
and
\begin{equation}
\frac{d\Phi(r)}{dr}=\frac{G M(r)}{r^2}
\label{eq:ness3}
\end{equation}
used to find the equilibrium structure of a spherically symmetric body of isotropic material when general-relativistic corrections are not important. Here $\Phi(r)$ is the Newtonian potential of the gravitational field inside the star.

%%%%%%%%%%%%%%%%%%%%%%%%%%%%%%%%%%%%%%%%%%%%%%%%%%%%%%%%%%%%%%%%%%%%%%%%%%%%%%%%%%%%%%%%%%%%%%%%%%%%%%%%%%%%%%%%%%%%%%%%%%%%%%%%%%%%%%%%%%%%%%%%
\section{Results}\label{sec:3}

By solving the TOV equations Eqs.~(\ref{eq:nu})-(\ref{eq:pr}) numerically, we obtained mass-central density $M-\rho$, and mass-radius $M-R$ relations for static hot WDs.

\begin{figure}[t]
\centerline{\includegraphics[scale=0.58]{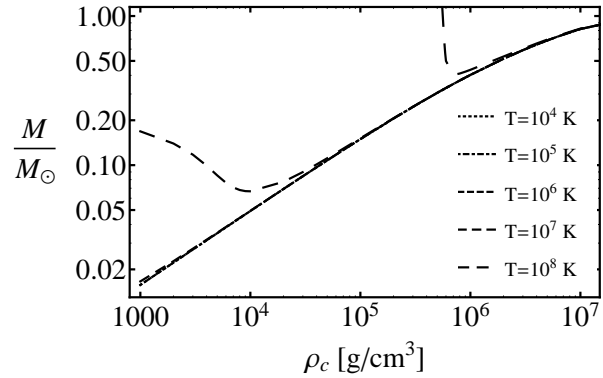}}
\caption{Total mass versus central density for $\mu=2$ white dwarfs for selected temperatures from $T=10^4$ K to  $T=10^8$ K.}\label{fig:f4}
\end{figure}

In Fig.~\ref{fig:f4} we plot the mass-central density for $\mu=2$ white dwarfs using the Chandrasekhar EoS at different temperatures in the range of densities where the finite-temperature effects are more important.

\begin{figure}[t]
\centerline{\includegraphics[scale=0.58]{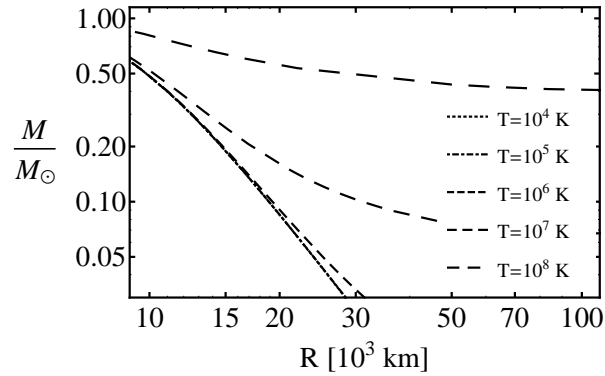}}
\caption{Mass versus radius for $\mu=2$ white dwarfs at temperatures $T=[\ 10^4, 10^5, 10^6, 10^7, 10^8]\ $K in the range $R=(10^4-10^5)$ km.}\label{fig:f3}
\end{figure}

In addition, in Fig.~\ref{fig:f3} we show the mass-radius relations. One can see that the finite-temperature effects are essential for small mass WDs. 

\begin{figure}[!hbtp]
\centerline{\includegraphics[scale=0.58]{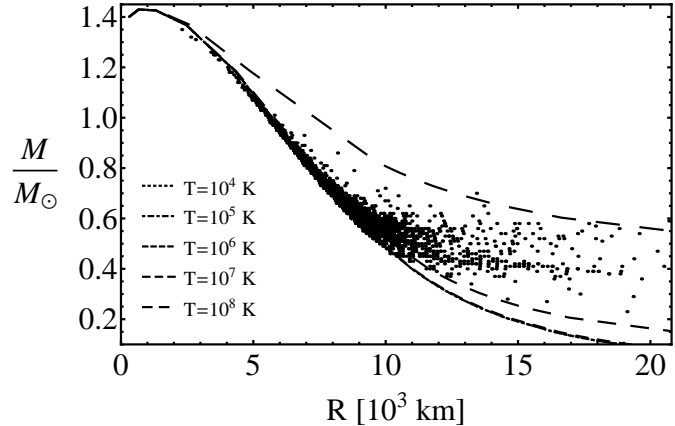}}
\caption{Mass-radius relations of white dwarfs obtained with the Chandrasekhar EoS (dashed black lines) for selected finite temperatures from $T=10^4$~K to $T=10^8$ K and their comparison with the masses and radii of white dwarfs taken from the Sloan Digital Sky Survey Data Release 4 (black dots).}\label{fig:f2}
\end{figure}

In Fig.~\ref{fig:f2} we plot $M-R$ relation for hot static white dwarfs with $\mu=2$ (see Fig.~\ref{fig:f4}) and compare them with the estimated masses and radii of white dwarfs from the Sloan Digital Sky Survey Data Release 4 (see Ref.~\refcite{tremblay}). 

Note, the temperature we use here is the temperature of the WD isothermal core $T_c$. In order to find the surface effective temperature $T_{eff}$, one needs to know the temperature gradient (so the heat flux) between the isothermal core and the surface of the star from which photons finally escape, i.e. a $T_c$-$T_{eff}$ relation. We adopt here the Koester formula  $T_{eff}^4/g=2.05\times10^{-10} T_c^{2.56}$, where $g$ is the surface gravity.\cite{koester2}

%%%%%%%%%%%%%%%%%%%%%%%%%%%%%%%%%%%%%%%%%%%%%%%%%%%%%%%%%%%%%%%%%%%%%%%%%%%%%%%%%%%%%%%%%%%%%%%%%%%%%%%%%%%%%%%%%%%%%%%%%%%%%%%%%%%%%%%%%%%%%%%%
\section{Conclusion}\label{sec:6}
In this work we studied the properties of static WDs by using the Chandrasekhar EoS at finite-temperatures. To investigate WDs in GR we solved the TOV equation numerically in order to construct $M-\rho$ and $M-R$  relations at finite temperatures. Furthermore we superposed our results with the estimated values of masses and radii obtained by Tremblay et al.\cite{tremblay} As a result we showed that all observational data, at least in the range of low masses, can be described by the Chandrasekhar EoS at finite-temperatures.
In our computations we used the values of temperature of the core of the WD. To compare with the real surface temperature of WDs we exploited the Koester formula which establishes the connection between the effective surface temperature and the temperature of the core of WDs. We found that most of the observed WDs have core temperatures lower than $10^8$~K  (see Fig. ~\ref{fig:f2}). A precise analysis using empirical mass-radius relations obtained from the spectroscopic or photometric measurements of masses and radii is still needed to confirm and extend our results.
%%%%%%%%%%%%%%%%%%%%%%%%%%%%%%%%%%%%%%%%%%%%%%%%%%%%%%%%%%%%%%%%%%%%%%%%%%%%%%%%%%%%%%%%%%%%%%%%%%%%%%%%%%%%%%%%%%%%%%%%%%%%%%%%%%%%%%%%%%%%%%%%
\section*{Acknowledgments}

%We would like to thank ... for your review and advice of our manuscript. We are grateful to the .... for the helpful comments and discussions (suggestions). 

This work was supported by the Ministry of Education and Science of the Republic of Kazakhstan, Grant No. 3101/GF4 IPC-11/2015.

\end{document}